\documentclass[prb,showpacs,twocolumn,preprintnumbers,amsmath,amssymb,floatfix]{revtex4}
\usepackage[dvips]{graphicx}
\usepackage[latin1]{inputenc}
\begin{document}
\draft

\hyphenation{a-long}

\title{Superconducting phase fluctuations in SmFeAsO$_{0.8}$F$_{0.2}$\\
from diamagnetism at low magnetic field above $T_{c}$}

\author{G. Prando,$^{1,2}$ A. Lascialfari,$^{1,3,4}$ A. Rigamonti,$^{1}$
L. Roman\'o,$^{5}$ S. Sanna,$^{1}$ M. Putti,$^{6}$ M.
Tropeano$^{6,7}$}

\address{$^{1}$Department of Physics ``A. Volta,'' University of Pavia-CNISM, I-27100 Pavia, Italy}
\address{$^{2}$Department of Physics ``E. Amaldi,'' University of Roma Tre-CNISM, I-00146 Roma, Italy}
\address{$^{3}$Department of Molecular Sciences Applied to Biosystems, University of Milano, I-20134 Milano, Italy}
\address{$^{4}$Centro S3, CNR-Istituto di Nanoscienze, I-41125 Modena, Italy}
\address{$^{5}$Department of Physics, University of Parma-CNISM, I-43124 Parma, Italy}
\address{$^{6}$CNR-SPIN and University of Genova, I-16146 Genova, Italy}
\address{$^{7}$Columbus Superconductors SpA, I-16133 Genova, Italy}

\widetext

\begin{abstract}
Superconducting fluctuations (SF) in SmFeAsO$_{0.8}$F$_{0.2}$
(characterized by superconducting transition temperature $T_{c}
\simeq 52.3$ K) are investigated by means of isothermal
high-resolution dc magnetization measurements. The diamagnetic
response above $T_{c}$ to magnetic fields up to $1$ T is similar to
that previously reported for underdoped cuprate superconductors and
justified in terms of metastable superconducting islands of non-zero
order parameter lacking long-range coherence because of strong phase
fluctuations. In the high-field regime ($H \gtrsim 1.5$ T) scaling
arguments predicted on the basis of the Ginzburg-Landau theory for
conventional SF are confirmed, at variance with what is observed in
the low-field regime. This fact shows that two different phenomena
are simultaneously present in the fluctuating diamagnetism, namely
the phase SF of novel character and the conventional SF. High
magnetic fields ($1.5$ T $\lesssim H \ll H_{c2}$) are found to
suppress the former while leaving unaltered the latter.
\end{abstract}


\maketitle

\narrowtext

\section{Introduction}

High-temperature superconductivity in Fe-based oxy-pnictides has
aroused strong interest among condensed matter physicists after its
discovery in 2008.\cite{Kam08} In spite of the huge amount of both
theoretical and experimental activities\cite{Lum10,Joh10} several
questions are still open, particularly regarding the pairing
mechanism. The symmetry of the order parameter has been investigated
by means of Josephson tunneling and ARPES experiments. Results show
that in Fe-based superconductors (both in doped REFeAsO, where RE
stands for rare-earth ion, and doped (Ba,Sr)Fe$_{2}$As$_{2}$
compounds, belonging to the so-called 1111 and 122 families
respectively) $s$-wave singlet superconductivity is at
work.\cite{Che10,Nak09} A multi-band scenario has been proposed,
where the appearance of two different superconducting gaps seems to
characterize Fe-based pnictide materials. BCS theory anyway cannot
account for either the high values of $T_{c}$ or the temperature
dependence of the superconducting gaps, similarly to the case of
MgB$_{2}$.\cite{Iav02} Eliashberg theory for strong-coupling
resulting in an interband $s^{\pm}$-wave model\cite{Maz08,Umm09} is
needed in order to account for the experimental results from
Josephson tunneling on NdFeAsO$_{0.88}$F$_{0.12}$\cite{Che10} and
from point-contact Andreev-reflection spectroscopy on
SmFeAsO$_{1-x}$F$_{x}$.\cite{Dag09}

Besides these similarities with $s$-wave superconductors and with
MgB$_{2}$, many experimental evidences emphasize affinities of
Fe-based oxy-pnictides with the cuprate high-temperature
superconductors (HTSC). By means of magnetoresistivity\cite{Lee10}
and ac susceptibility\cite{Pra11} it has been pointed out that the
features of vortex dynamics in SmFeAsO$_{1-x}$F$_{x}$ can be mapped
onto models successfully applied to cuprate materials,\cite{Bla94}
as, for instance, the thermally-activated flux flow of vortex lines.
Moreover, scaling arguments from vortex-glass theory developed for
$I-V$ characteristics of cuprates\cite{Koc89,Fis91} were found to
describe experimental data also in SmFeAsO$_{0.85}$.\cite{Lee10} The
existence of a pseudo-gap precursor state also in
SmFeAsO$_{1-x}$F$_{x}$, finally, has been suggested by different
experimental evidences.\cite{Che08,Sat08,Mer09,Gon09,Pal11}

Small coherence lengths, reduced carriers density, high transition
temperatures and marked anisotropy are all factors causing a strong
enhancement of superconducting fluctuations
(SF).\cite{Sko75,Tin96,Lar05} The generation of fluctuating Cooper
pairs above $T_{c}$ results in the appearance of a Langevin-type
diamagnetic contribution to the magnetization $- M_{dia}(T,H)$
existing side by side with the paramagnetic contribution from
fermionic carriers. Since the size $\xi(T)$ of the fluctuating pairs
grows when $T$ approaches the transition temperature $T_{c}$ from
above, $\left|M_{dia}(T,H)\right|$ should diverge near the
transition for any small fixed magnetic field, being zero for $H =
0$ Oe. On the other hand, it is evident that very strong magnetic
fields, comparable to $H_{c2}(0)$, must suppress SF. Thus, the
isothermal magnetization curve $- M_{dia}(\overline{T},H)$
($\overline{T}$ is a fixed temperature) has to exhibit an upturn.
The value of the upturn field $H_{up}$ in the magnetization curves
can approximately be considered inversely proportional to the
coherence length.\cite{Sko75,Tin96,Lar05} Thus, in optimally doped
cuprates, $H_{up}$ is expected at very strong magnetic fields. At
variance, in underdoped superconducting cuprates, the magnetization
curves above $T_{c}$ evidence an upturn for $H \ll
H_{c2}(0)$.\cite{Las02a,Las02b}

As shown in the following, similar effects have been found in
Fe-based oxy-pnictides belonging to the REFeAsO$_{1-x}$F$_{x}$
family, evidencing a further analogy with cuprate superconductors.
SF in Fe-based oxy-pnictides have already been theoretically
discussed.\cite{Mur10} Calorimetric measurements performed on
SmFeAsO$_{0.85}$F$_{0.15}$ single-crystals have been recently
interpreted in terms of Ginzburg-Landau SF.\cite{Wel11} Furthermore,
a magnetic investigation of Ba$_{1-x}$K$_{x}$Fe$_{2}$As$_{2}$
single-crystals has been carried out in the high-field
regime\cite{Mos11} and classical Ginzburg-Landau scaling has been
observed. Until now, to our knowledge, no claim of non-classical SF
in Fe-based oxy-pnictide superconductors has been reported yet.

This paper deals with high-resolution magnetization measurements
above $T_{c}$ in SmFeAsO$_{0.8}$F$_{0.2}$. The observed
phenomenology at low magnetic fields $H \lesssim 1$ T can be
well-described in terms of the superconducting phase fluctuation
model, confirming a analogy of SmFeAsO$_{0.8}$F$_{0.2}$ with
underdoped cuprate superconductors. In this picture,
extra-diamagnetism arises above $T_{c}$ due to the appearance of
mesoscopic superconducting ``islands'' with nonzero order parameter
at frozen amplitude lacking of coherence due to marked phase
fluctuations.\cite{Sew01,Rom03} In the high-field range $1.5$ T
$\lesssim H \ll H_{c2}$ the suppression of superconducting phase
fluctuations is found to leave unaltered the classical
Ginzburg-Landau fluctuations, making scaling arguments applicable.

\section{Experimental}

Powders of SmFeAsO$_{0.8}$F$_{0.2}$ were prepared by solid state
reaction at ambient pressure from Sm, As, Fe, Fe$_{2}$O$_{3}$ and
FeF$_{2}$.\cite{Mar08} SmAs was first synthesized from pure elements
in an evacuated, sealed glass tube at a maximum temperature of
550°C. The final sample was synthesized by mixing SmAs, Fe,
Fe$_{2}$O$_{3}$ and FeF$_{2}$ powders in stoichiometric proportions,
using uniaxial pressing to obtain pellets and then thermal treating
in an evacuated, sealed quartz tube at 1000°C for 24 hours, followed
by furnace cooling. The sample was analyzed by powder X-ray
diffraction in a Guinier camera, with Si as internal standard. The
powder pattern showed the sample to be nearly single phase with two
weak extra lines at low angle attributable to SmOF. The lattice
parameters $a = 3.930(1)$ Å and $c = 8.468(2)$ Å have been derived,
in agreement with data reported elsewhere.\cite{Mar09} The magnetic
characterization of the sample was reported in a previous paper,
together with a {}$^{19}$F-NMR investigation allowing one to infer
that the pairing mechanism is uncorrelated with fluctuating
Sm$^{3+}$ magnetic moments.\cite{Pra10}

The magnetization $M$ measurements were carried out by means of
Quantum Design MPMS-XL7 SQUID magnetometer. The experimental data
were collected both in zero-field-cooled (ZFC) and field-cooled (FC)
conditions. In ZFC measurements, the magnetic field is applied at a
given temperature after cooling the sample across the
superconducting critical temperature $T_{c}$ in zero applied field.
In FC measurements the magnetic field is applied
at temperatures $T \gg T_{c}$.

\section{Superconducting fluctuations vs. precursor
diamagnetism}\label{SectSF}

The response of superconducting materials to the application of
magnetic fields at $T \gtrsim T_{c}$ shows a diamagnetic
contribution to the magnetization ($M_{dia}$) superimposed on the
Pauli-like paramagnetic term due to fermionic
carriers.\cite{Lar05,Tin96} This effect may be due to different
phenomenologies and could arise from different origins, as explained
in detail below.

\subsection{Ginzburg-Landau superconducting fluctuations}

In most cases, as in conventional metallic superconductors, cuprate
materials and MgB$_{2}$, the appearance of a diamagnetic term at $T
\gtrsim T_{c}$ can be associated with fluctuations of the order
parameter $\psi = \left|\psi\right| e^{\imath \theta}$
(superconducting fluctuations,
SF).\cite{Kos94,Hub95,Mis00,Las02a,Rom05,Ber08} In particular, in
the classical Ginzburg-Landau framework of second-order phase
transitions, one has to take into account fluctuations of the
amplitude of $\psi$ around its equilibrium value $\left|\psi\right|
= 0$, leading the quantity
$\sqrt{\langle\left|\psi^{2}\right|\rangle}$ to assume non-zero
values for $T \gtrsim T_{c}$.\cite{Lar05,Tin96}

The Ginzburg-Landau picture must be corrected to take into account
the field effects on the Cooper pairs and the anisotropic
free-energy functional. In three-dimensional (3D) superconductors
for $T \simeq T_{c}$ one has $M_{dia} \propto \sqrt{H}$ (Prange
regime)\cite{Lar05,Tin96} and by increasing the field saturation of
$\left|M_{dia}\right|$ is expected. Furthermore, isothermal
$M_{dia}/\sqrt{H}$ vs. $H$ curves are expected to cross at $T_{c} (H
= 0)$.\cite{Kos94,Jun98}

In optimally-doped cuprate superconductors, the fluctuating
diamagnetic contribution is expected to be suppressed by the
application of magnetic fields $H \sim H_{c2}$. This corresponds to
the appearance of an upturn field $H_{up}$ defined as the value of
magnetic field at which $\left|M_{dia}\right|$ starts to decrease on
increasing $H$, as already mentioned. For optimally-doped cuprate
HTSC, the upturn field could occur at very high values, typically of
the order of tens of Teslas. In fact, $H_{up}$ can be crudely
justified by assuming a first order correction in which the
fluctuations-induced evanescent superconducting droplets are
spherical, with a diameter $d$ of the order of the coherence length
$\xi(T)$. If $d \ll \xi(T)$, the zero-dimensional approximation can
be used\cite{Ber06} where the order parameter is no longer dependent
on spatial degrees of freedom. In this approach, the value of the
upturn field $H_{up}$ becomes inversely proportional to the square
of the coherence length $\xi(T \rightarrow 0)$: $H_{up} = \epsilon
\Phi_{0}/\xi^{2}(0)$, where $\epsilon = \left(T -
T_{c}\right)/T_{c}$ and $\Phi_{0} = hc/2e = 2.0679 \cdot 10^{-7}$ G
cm$^{2}$ corresponds to the flux quantum. In conventional BCS
superconductors or in MgB$_{2}$, $H_{up}$ is rather small (typically
around $50$ to $100$ Oe) while in HTSC the upturn in $M_{dia}$ vs.
$H$ could be detected only at very high fields and in optimally
doped YBa$_{2}$Cu$_{3}$O$_{6+\delta}$ (YBCO) no upturn in $M_{dia}$
has been observed up to 7 Teslas.\cite{Jun98,Las02b,Las03a,Las03b}

\subsection{Phase fluctuations and precursor diamagnetism}

At variance with the findings recalled above, underdoped cuprate
materials display much richer phenomenologies than the optimally
doped ones.\cite{Las02b,Las03a,Las03b,Li10,Ber10} In particular,
upturn fields $H_{up} \simeq 10 - 1000$ Oe have been evidenced. This
phenomenon can be justified by taking into account the fluctuations
of the phase of the superconducting order parameter in mesoscopic
``islands'', the long-range superconductivity being prevented by the
lack of coherence due to marked phase
fluctuations.\cite{Sew01,Rom03,Eme95} The existence of these islands
is supported by scanning microscopy, at least in
La$_{2-x}$Sr$_{x}$CuO$_{4}$ (LSCO) and in Bi-based
cuprates,\cite{Xu00,Igu01,Lan02} and by the detection of a large
Nernst signal above $T_{c}$.\cite{Wan06} The fluctuations of the
order parameter phase imply the presence of thermally induced
vortices which add to those generated by the magnetic field.

The starting point to derive the magnetic susceptibility is the
Lawrence-Doniach-like functional for a layered system:\cite{Tin96}
\begin{widetext}
\begin{equation}\label{EqLDFunctional}
    \mathcal{F}_{LD} \left[\theta\right] = \frac{1}{s} \sum_{l} \int
    d^{2}\textbf{r} \left\{J_{\parallel} \left(\nabla_{\parallel}
    \theta - \frac{\imath 2e}{\hbar c} A_{\parallel}\right)^{2} +
    J_{\perp}\left[1 - \cos\left(\theta_{l+1} - \theta_{l}\right)
    \right]\right\}
\end{equation}
where $\left|\psi\right|^{2}$ is frozen at a constant value and only
the dependence on the phase $\theta$ is considered. In Eq.
\ref{EqLDFunctional} the index $l$ labels the different
superconducting layers (separated by a distance $s$),
$J_{\parallel}$ and $J_{\perp}$ are the order parameter phase
coupling constants within the layers and between different layers
(respectively) and the potential vector $\textbf{A}_{\parallel}$
describes both the magnetic field applied parallel to the $c$ axis
and the one induced by thermal fluctuations. The second derivative
of the free energy resulting from Eq. \ref{EqLDFunctional} yields
the field dependent susceptibility $\chi(H)$:
\begin{equation}\label{EqChiVsH}
    \chi (H) = -\frac{k_{B}T}{\Phi_{0}^{2}}\frac{1}{s \left(1 +
    2n\right)} \left\{\frac{\left[1 + \left(\frac{H}{H^{*}}\right)^{2}
    \delta\right]^{2}}{n_{vor}} + s^{2}\gamma^{2} \left(1 + n\right)
    n \left[1 + \left(\frac{H}{H^{*}}\right)^{2} \delta\right]^{2}\right\}
    + \frac{47\pi L^{2}}{540} \frac{J_{\parallel}}{s} \left(\frac{2\pi}
    {\Phi_{0}}\right)^{2} \left(\frac{H}{H^{*}}\right)^{2} \delta
\end{equation}
\end{widetext}
where $\delta = J_{\parallel}/k_{B}T$, $H^{*} \equiv \Phi_{0}/L^{2}$
is a characteristic field related to the size $L$ of the islands,
while $\gamma \equiv \xi_{ab} (0)/ \xi_{c} (0)$ is the anisotropy
factor. In Eq. \ref{EqChiVsH}, $n_{vor}= n_{H} + n_{th}$ is the
total vortex density due to the field induced vortices $n_{H} =
H/\Phi_{0}$ and the thermally excited vortices, whose density
$n_{th}$ is affected by the applied field also depending on the
number $n$ of correlated layers:
\begin{equation}\label{EqNumbVort}
    n_{th} = n_{0} \exp\left\{-\frac{E_{0}\left(1 + 2n\right)}
    {k_{B}T\left[1 + \left(\frac{H}{H^{*}}\right)^{2} \delta\right]}
    \right\}.
\end{equation}

When faced with values of upturn fields $H_{up}$ in the order of
$100$ to $1000$ Oe one has to be careful in the data interpretation,
since other mechanisms could possibly lead to the observed
phenomenology. In disordered systems such as in Al-doped MgB$_{2}$
or YNi$_{2}$B$_{2}$C, for instance, the average bulk $T_{c}$ arises
from a spatial distribution of local transition temperatures
$T_{c}(\textbf{r})$ inside the sample.\cite{Ber08,Cab06} The main
effect of inhomogeneity is to induce a precursor diamagnetism above
$T_{c}$ associated with the diamagnetic response of those regions
where $T_{c}(\textbf{r}) > T_{c}$ holds locally. Also in this case,
high magnetic fields tend to reduce the extra-diamagnetic response
above $T_{c}$ giving rise to an upturn field in the isothermal
magnetization curves resulting from a distribution of critical
fields $H_{c1}(\textbf{r})$ associated to the spatial regions where
$T_{c}(\textbf{r}) > T_{c}$. In this case, $H_{up}$ mimics the
behaviour of $H_{c1}(T)$.

A straightforward way of distinguishing extra-diamagnetism due to
phase fluctuations from precursor diamagnetism due to sample
inhomogeneity lies in the different temperature dependence of the
upturn field in the two regimes.\cite{Rig05} $H_{up}$, in fact,
increases on increasing temperature in the phase SF scenario while
it decreases in disorder-induced diamagnetism.

\section{Experimental results}\label{SectResults}

A detailed analysis of the different contributions to the
macroscopic magnetization $M$ of the sample together with the
procedure to extract $M_{dia}$ from the isothermal magnetization
curves will be outlined in the following.

\begin{figure}[htbp]
\vspace{6.6cm} \includegraphics{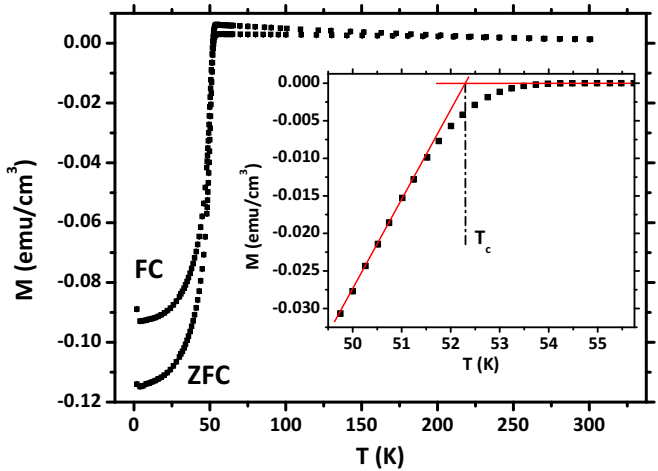} \caption{\label{FigSCTrans}(Color online)
$M$ vs. $T$ magnetization curves in both ZFC and FC conditions at $H
= 5$ Oe. Inset: blow up of the onset region in the FC curve after
subtracting a linear contribution accounting for other sources of
magnetism in this narrow $T$-range (see text). The superconducting
transition temperature can be estimated as $T_{c} \simeq 52.3$ K.}
\end{figure}

$M$ vs. $T$ curves at small magnetic field ($H = 5$ Oe) obtained
both in ZFC and FC conditions are reported in Fig. \ref{FigSCTrans}.
A slight separation between FC and ZFC data already at $T \gg T_{c}$
implies the presence of a small amount of spurious magnetic phases.
The transition temperature $T_{c} (H \rightarrow 0) = \left(52.3 \pm
0.1\right)$ K can be estimated by first subtracting in a few-K
region around the diamagnetic onset a contribution associated with
that spurious magnetic signal (see details subsequently). Then,
$T_{c}$ is evaluated as the intersection of two linear fits of the
resulting data (well below and well above $T_{c}$, as shown in Fig.
\ref{FigSCTrans}, inset).

\begin{figure}[htbp]
\vspace{6.6cm} \includegraphics{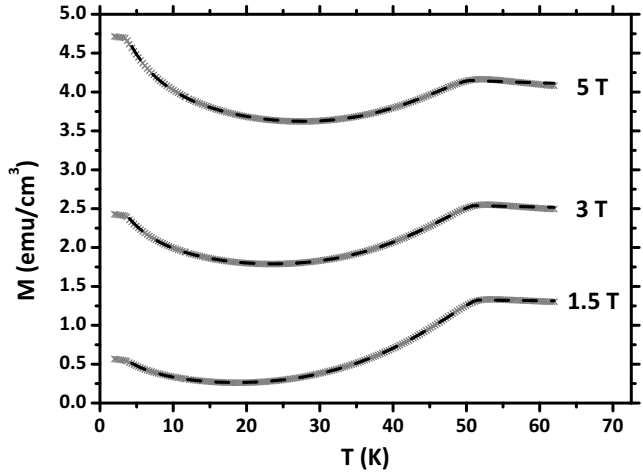} \caption{\label{FigMvsT}$M$ vs. $T$
magnetization curves in FC conditions at different magnetic fields.
The data are fitted according to a phenomenological expression (Eq.
\ref{EqScTrans}, see dashed lines). Low-temperature anomalies ($T
\simeq 4$ K) are related to the antiferromagnetic ordering of
Sm$^{3+}$ magnetic moments.\cite{Rya09}}
\end{figure}

In Fig. \ref{FigMvsT} representative raw $M$ vs. $T$ curves at high
magnetic fields are shown. The data are well fitted by the
phenomenological function
\begin{eqnarray}\label{EqScTrans}
    M_{dc}(T) & = & M_{sc} \left[1 -
    \left(\frac{T}{T_{c}}\right)^{\alpha}\right]^{\beta}
    +\nonumber\\
    & & + C_{cw}\frac{H}{T - T_{N}} + M_{0}.
\end{eqnarray}
The first term is the diamagnetic Meissner response (which we have
empirically fitted by a two-exponents mean field function). The
second term in Eq. \ref{EqScTrans} is the Curie-Weiss paramagnetic
contribution associated with the Sm$^{3+}$ sublattice, with $T_{N} =
4$ K as fixed parameter. The best fit of the Curie-Weiss constant
$C_{cw}$ leads to a magnetic moment $\mu = \left(0.32 \pm
0.01\right)$ $\mu_{B}$ per Sm$^{3+}$ ion. This is in close agreement
with the value $\mu \simeq 0.53$ $\mu_{B}$ reported from a
neutron-diffraction experiment in the antiferromagnetically-ordered
phase.\cite{Rya09} The tetragonal crystalline electrostatic
environment is known to be described in the Stevens' operators
formalism as
\begin{equation}
    \hat{\mathcal{H}}_{CEF} = B_{2}^{0}\hat{O}_{2}^{0} +
    B_{4}^{0}\hat{O}_{4}^{0} + B_{4}^{4}\hat{O}_{4}^{4}
\end{equation}
leading in the case of both Sm$^{3+}$ and Ce$^{3+}$ to a splitting
of the ground state $J = \frac{5}{2}$ multiplet into three
doublets.\cite{Van32,Chi08} However, crystal-field effects on the
magnetization can be safely disregarded in the examined temperature
range ($T < 60$ K), thus justifying the use of a simple Curie-Weiss
fitting function for the paramagnetic
contribution.\cite{Van32,Chi08,Bak09,Cim09} Finally, the last
$T$-independent term in Eq. \ref{EqScTrans} arises from different
contributions including a Pauli-like susceptibility associated with
itinerant electrons, ionic diamagnetism and a small amount of
ferromagnetic impurities. The impurity contribution $M_{imp} \simeq
0.23$ emu/cm$^{3}$ (saturated value for $H \gtrsim 1$ T) was
quantified from a isothermal $M$ vs. $H$ curve at $T = 56$ K and,
more precisely, from the intercept value of the linear fit of the
paramagnetic contribution at $H = 0$ Oe. The parameters obtained
from the fitting procedure at different magnetic fields in the range
$1.5$ T $\leq H \leq 7$ T have been reported in Tab. \ref{TabFit}.

\begin{table}[htbp]
\caption{\label{TabFit} Parameters associated with the fitting
procedure of $M$ vs. $T$ curves (see Fig. \ref{FigMvsT} for
representative raw data) at different $H$ values (see first column,
units: T) by means of the phenomenological expression reported in
Eq. \ref{EqScTrans}. $M_{sc}/H$ quantifies the amplitude of the
diamagnetic response from the superconducting phase (units:
erg/Oe$^{2}$ cm$^{3}$), $T_{c}$ is the superconducting critical
temperature (units: K), $\alpha$ and $\beta$ are the two exponents
of the mean-field-like phenomenological contribution representing
the superconducting shielding, $C_{cw}$ is the Curie-Weiss constant
associated with the paramagnetic phase of the Sm$^{3+}$ sublattice
(units: erg K/Oe$^{2}$ cm$^{3}$) and $M_{0}/H$ represents all the
other sources of $T$-independent magnetism (units: erg/Oe$^{2}$
cm$^{3}$).}
\begin{ruledtabular}
\begin{tabular}{c|cccccc}%
$H$ & $10^{5} M_{sc}/H$ & $T_{c}$ & $\alpha$ &
$\beta$ & $10^{5} C_{cw}$ & $10^{5} M_{0}/H$\\
\toprule%
$1.5$ & $-7.95$ & $51.28$ & $3.54$ & $1.19$ & $25.26$ &
$8.34$\\
$2$ & $-5.42$ & $51.145$ & $3.68$ & $1.22$ & $24.66$ &
$8.13$\\
$2.5$ & $-3.98$ & $51.05$ & $3.82$ & $1.25$ & $24.39$ &
$8.10$\\
$3$ & $-3.11$ & $50.96$ & $3.97$ & $1.29$ & $24.09$ &
$8.02$\\
$3.5$ & $-2.49$ & $50.89$ & $4.09$ & $1.33$ & $23.93$ &
$8.02$\\
$4$ & $-2.07$ & $50.815$ & $4.24$ & $1.37$ & $23.68$ &
$7.97$\\
$5$ & $-1.51$ & $50.675$ & $4.47$ & $1.44$ & $23.33$ &
$7.88$\\
$6$ & $-1.16$ & $50.545$ & $4.69$ & $1.51$ & $23.01$ &
$7.75$\\
$7$ & $-0.93$ & $50.43$ & $4.94$ & $1.59$ & $22.76$ &
$7.62$\\
\end{tabular}
\end{ruledtabular}
\end{table}

In order to obtain reliable estimates of the SF diamagnetic
contribution $M_{dia}$, one has to allow for all the additional
contributions discussed above. At this aim, we employed the
subtraction procedure successfully used in the analysis of cuprate
materials and of MgB$_{2}$. Isothermal $M$ vs. $H$ curves ($H < 5$
T) were measured at selected temperatures around the superconducting
onset ($50$ K $< T < 56$ K) and the high-field linear paramagnetic
contribution was individually subtracted from those curves to
account for its $T$-dependence. The so-called reference isothermal
curve (relative, in this case, to $T = 56$ K) was then subtracted to
each obtained curve. By doing this, one subtracts all the other
sources of $T$-independent magnetism, by considering the fact that
the spurious contribution is practically constant in this narrow $T$
range. The reference $T = 56$ K was chosen by carefully examining
the onset region in Fig. \ref{FigSCTrans}. In particular, a value
close enough to the investigated region was selected to neglect the
possible $T$-dependence of spurious contributions, meantime safely
far from the fluctuative region itself.

\begin{figure}[htbp]
\vspace{6.6cm} \includegraphics{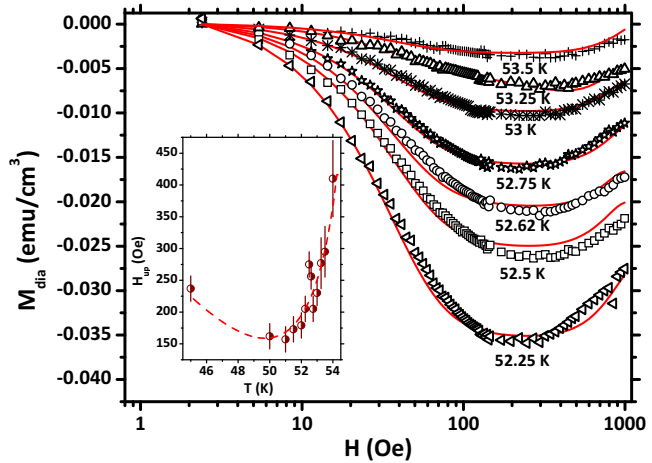} \caption{\label{FigIsotFL}(Color online)
Isothermal diamagnetic contributions $M_{dia}$  vs. $H$ for
representative temperatures above $T_{c}$, displaying upturn fields
in the range $H_{up} \sim 200$ to $400$ Oe. The curves have been
obtained after the subtraction procedure described in the text.
Continuous lines are best-fits obtained by means of numerical
integration of Eq. \ref{EqChiVsH}. Inset: temperature dependence of
the upturn field, clearly increasing on increasing $T$ above
$T_{c}$. The dashed line is a guide for the eye.}
\end{figure}

In Fig. \ref{FigIsotFL} representative isothermal magnetization
curves $M_{dia}$ vs. $H$ obtained in ZFC conditions at different
temperatures above $T_{c} (H = 0)$ are shown in the low-field range
(up to $H \simeq 1000$ Oe). From the analysis of the data
interesting insights are obtained. One notices the appearance of
upturn fields in the order of $200$ to $400$ Oe. More interestingly,
it is noted (see Fig. \ref{FigIsotFL}, inset) that, in the
temperature limit $T
> T_{c}$, $H_{up}$ increases on increasing temperature. As
emphasized in Sect. \ref{SectSF}, both the small values of $H_{up}$
and the increase of $H_{up}$ on increasing temperature suggest that
the observed diamagnetism above $T_{c}$ can be ascribed to the
presence of phase SF. The effects of the $T_{c} (\textbf{r})$
distribution, due to some sample inhomogeneity, can safely be
neglected. This does not imply that the sample is totally
homogeneous but rather that the phase fluctuations play a dominant
role in the diamagnetism above $T_{c}$ at the examined fields.

In SmFeAsO$_{0.8}$F$_{0.2}$ the value of the coherence length is
still controversial. By examining paraconductivity data, analyzed
with a four-band three-dimensional model, one derives $\xi_{0} = 19$
Å,\cite{Fan09} while using a two-dimensional lowest Landau level
scaling and $H_{c2}$ measurements in single-crystal samples the same
quantity ranges from $3$ to $4$ Å.\cite{Pal09,Lee09} Thus, the
estimated values for $\xi_{0}$ are anyway small, in analogy with
those found in cuprate materials. Then, in the framework of
classical GL fluctuations for $\epsilon \sim 10^{-2}$, $H_{up}$
would be expected to vary between 5 T and 25 T. It is evident that
the occurrence of an upturn field of the order of few hundred Gauss
in the magnetization curves of Fig. \ref{FigIsotFL} cannot be
ascribed to the saturation of magnetization expected in GL
fluctuations regime.

\section{Discussion and conclusions}

By means of numerical integration of Eq. \ref{EqChiVsH}, the
magnetization curves as a function of field are obtained (solid
lines in Fig. \ref{FigIsotFL}). The parameters corresponding to the
best fit are collected in Tab. \ref{Tab}. The interlayer distance
has been set $s = 8.5$ Å.

\begin{table}[htbp]
\caption{\label{Tab} Values of $H^{*}$ and of the size of the
superconducting islands $L$ resulting from the best-fit procedures.
The parameter $N$ should be considered an order-of-magnitude
estimate of the number of islands having assumed superconducting
regions of volume given by $L^{2} d$ with the value of the depth $d$
fixed to $25.5$ Å (corresponding to $n = 3$, see text).}
\begin{ruledtabular}
\begin{tabular}{ccccc}%
$T$ (K) & $H^{*}$ (Oe) & $L$ (nm) & $N$
(cm$^{-3}$)\\
\toprule%
$52.25$ & $883$ & $154$ & $2.4 \cdot 10^{14}$\\
$52.5$ & $888$ & $153$ & $1.5 \cdot 10^{14}$\\
$52.62$ & $1001$ & $145$ & $1.45 \cdot 10^{14}$\\
$52.75$ & $1101$ & $138$ & $1.25 \cdot 10^{14}$\\
$53$ & $1360$ & $124$ & $1 \cdot 10^{14}$\\
$53.25$ & $1440$ & $121$ & $5.2 \cdot 10^{13}$\\
$53.5$ & $1581$ & $115$ & $4.9 \cdot 10^{13}$\\
\end{tabular}
\end{ruledtabular}
\end{table}

According to the relation $n_{0} \simeq 10^{4}/a^{2}$, in Eq.
\ref{EqNumbVort} one has $n_{0} = 6.5 \cdot 10^{18}$ cm$^{-2}$, for
$a = 3.9$ Å.\cite{Mar09} The activation energy $E_{0}$ in Eq.
\ref{EqNumbVort} is usually estimated around $10$ $k_{B}T_{c}$ then
the term $E_{0} \left(1 + 2n\right)$, chosen as a free parameter,
allows an evaluation of $n$. In particular, $E_{0} \left(1 +
2n\right) = 28$ $k_{B}T_{c}$ indicating the number of correlated
layers $n = 3$, with an activation energy $E_{0} \simeq 4$
$k_{B}T_{c}$. Similar investigations in YBCO\cite{Las02b} and in
Sm-based cuprates\cite{Ber10} have been performed using $n = 3$, in
agreement with the fact that all these systems have similar
anisotropy parameters. For this reason in this analysis $\gamma = 7$
will be considered as a working hypothesis. Furthermore, $E_{0}
\left(1 + 2n\right)$ turns out almost temperature independent,
suggesting that the number of correlated layers does not appreciably
change in the temperature range that has been studied. In the
numerical integration procedure also $H^{*}$ and
$J_{\parallel}/k_{B}T$ are given as free parameters. From the
characteristic field $H^{*}$ it is possible to estimate an order of
magnitude of the average size $L$ of the superconducting regions
(see Tab. \ref{Tab}).

As might be expected, on increasing temperature the areas with
non-zero order parameter progressively reduce in size. The
progressive decrease of the volume occupied by the superconducting
regions are due both to the decrease in the average size of the
islands and/or to the decrease of their number.

The term $J_{\parallel}/k_{B}T$ results nearly independent on
temperature, around the value 2.5. In other situations, as in Sm
based cuprates,\cite{Ber10} this ratio decreases with increasing
temperature, according to its close relation to the superfluid
density, as suggested by the Berezinskii-Kosterlitz-Thouless
theory.\cite{Tin96} In the present case the fluctuation effect can
be detected in a temperature range too small to appreciate the
reduction.

\begin{figure}[htbp]
\vspace{6.6cm} \includegraphics{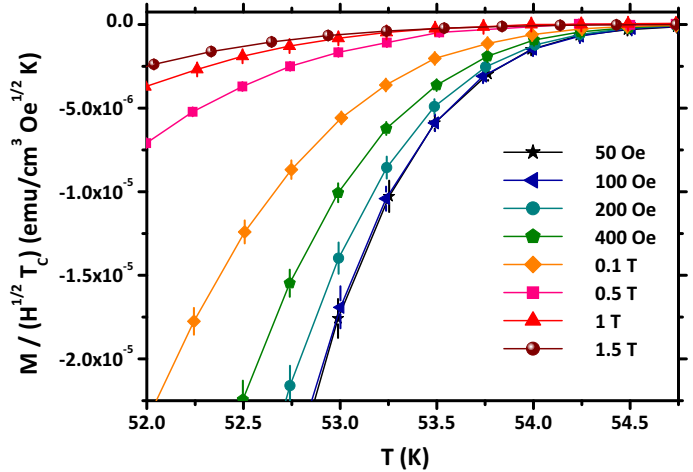} \caption{\label{FigNonCrossGL}(Color
online) Reduced magnetization $m_{c} = M_{dia}/\sqrt{H} T_{c}$ vs.
$T$ at magnetic fields $H \leq 1.5$ T. No crossing of curves at
$T_{c}$ is observed.}
\end{figure}

\begin{figure}[htbp]
\vspace{6.6cm} \includegraphics{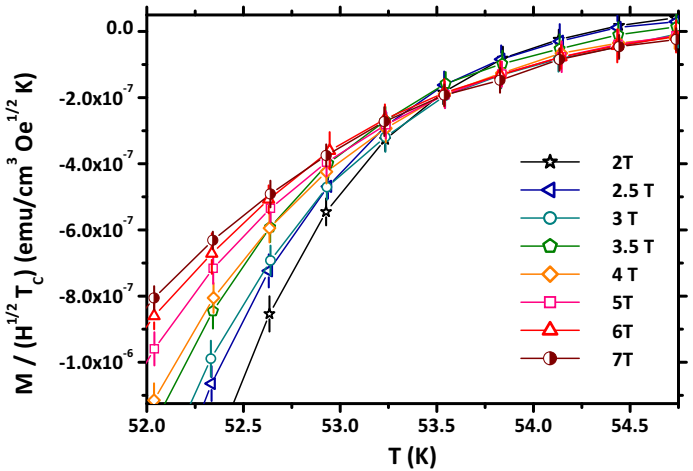} \caption{\label{FigCrossGL}(Color online)
Reduced magnetization $m_{c} = M_{dia}/\sqrt{H} T_{c}$ vs. $T$ at
magnetic fields $H \geq 2$ T. Crossing of curves at $T \simeq 53.2$
K suggests that the classical GL framework is applicable at high
fields, where the contribution from phase fluctuations is
suppressed.}
\end{figure}

A further confirmation of the inapplicability of the conventional GL
model, at least in the field range of Fig. \ref{FigIsotFL}, can be
drawn from analysis of the reduced magnetization $m_{c} =
M_{dia}/\sqrt{H} T_{c}$ vs. $T$, shown in Figs. \ref{FigNonCrossGL}
and \ref{FigCrossGL}. The data have been obtained by considering $M$
vs. $T$ curves at different magnetic fields in the range $0 < H < 7$
T (like those shown in Fig. \ref{FigMvsT}) and by individually
subtracting a linear contribution in a few-K region around the onset
to each curve accounting for all sources of magnetism other than
Meissner shielding and SF effects (as described in Sect.
\ref{SectResults}).

According to the GL theory for gaussian fluctuations and scaling
arguments for isotropic superconductors, $m_{c}$ should take the
universal value\cite{Tin96,Jun98,Sch93,Sch94}
\begin{equation}
    m_{c} = \frac{k_{B}}{\Phi_{0}^{3/2}} m_{3}(\infty) \simeq 4.6
    \cdot 10^{-7} \textrm{emu}/\textrm{cm}^{3} \textrm{Oe}^{1/2}
    \textrm{K}
\end{equation}
at $T \simeq T_{c}$ where the experimental data should cross. If the
system is anisotropic, the value of $m_{c}$ is expected to be
enhanced by a factor $\gamma$. In Fig. \ref{FigNonCrossGL} no
crossing in $m_{c}$ vs. $T$ data is observed at fields up to 1.5 T,
in contrast with the scaling prediction. Conversely, Fig.
\ref{FigCrossGL} shows that at higher fields, up to 7 T, the reduced
magnetization curves as a function of temperature cross at $T =
53.2$ K assuming the value $m_{c} \simeq 2.7 \cdot 10^{-7}$
emu/cm$^{3}$ Oe$^{1/2}$ K. Since the data refer to powders, the
value of $m_{c}$ must be multiplied by a factor 3, which takes into
account the powder average, as shown elsewhere in the case of
MgB$_{2}$.\cite{Las02a,Mis00} An anisotropy parameter $\gamma \sim
2$ would then be inferred. This value is smaller than the ones
reported in literature ($\gamma = 5$ to $9$) which however show
large uncertainty and appear to be related to the doping
level.\cite{Pal09,Lee09} Furthermore, the crossing temperature is
slightly above the critical temperature estimated from the inset of
Fig. \ref{FigSCTrans}. These aspects are possibly due to the
renormalization of the transition temperature due to the phase
fluctuations.\cite{Tal11} The effects related to the simultaneous
occurrence of non-conventional phase fluctuations and GL classical
fluctuations are worthy of further experimental and theoretical
studies. In conclusion, it appears that the fluctuation diamagnetism
above $T_{c}$ can be described by GL theory only at high fields,
while at a lower field range the phenomena highlighted in Fig.
\ref{FigIsotFL} require a different explanation based on the
occurrence of phase-fluctuations.

Summarizing, the diamagnetic response above $T_{c}$ at low magnetic
field in SmFeAsO$_{0.8}$F$_{0.2}$ cannot be described within the
classical Ginzburg-Landau approach. From this point of view, the
system is not comparable either to BCS superconductors or to
optimally-doped cuprates. On the contrary, the experimental results
support the picture based on the idea of precursor islands where the
amplitude of the order parameter is frozen, while the long-range
coherence associated with a bulk superconducting state is prevented
by marked fluctuations in the phase.

\begin{acknowledgements}
Stimulating discussions with P. Carretta and A. Varlamov are
gratefully acknowledged. A. Palenzona is acknowledged for the sample
preparation. One of us (G. P.) thanks E. Bernardi for an early
collaboration on fluctuating diamagnetism in metallic nanoparticles
and useful discussions concerning data analysis.

This work was partially supported by Compagnia di S. Paolo and by
MIUR under Project No. PRIN2008XWLWF9.
\end{acknowledgements}


\begin{references}
\bibitem{Kam08} Y. Kamihara, T. Watanabe, M. Hirano,
H. Hosono, {\it J. Am. Chem. Soc.} {\bf 130}, 3296 (2008)
\bibitem{Lum10} H. D. Lumsden and A. D. Christianson, {\it J. Phys.:
Cond. Matt.} {\bf 22}, 203203 (2010)
\bibitem{Joh10} D. C. Johnston, {\it Adv. Phys.} {\bf 59}, 803 (2010)
\bibitem{Che10} C.-T. Chen, C. C. Tsuei, M. B. Ketchen, Z.-A. Ren, Z. X.
Zhao, {\it Nature Phys.} {\bf 6}, 260 (2010)
\bibitem{Nak09} K. Nakayama, T. Sato, P. Richard, Y.-M. Xu, Y. Sekiba,
S. Souma, G. F. Chen J. L. Luo, N. L. Wang, H. Ding, T. Takahashi,
{\it Europhys. Lett.} {\bf 85}, 67002 (2009)
\bibitem{Iav02} M. Iavarone, G. Karapetrov, A. E. Koshelev, W. K. Kwok,
G. W. Crabtree, D. G. Hinks, W. N. Kang, E.-M. Choi, H.-J. Kim,
H.-J. Kim, S. I. Lee, {\it Phys. Rev. Lett.} {\bf 89}, 187002 (2002)
\bibitem{Maz08} I. I. Mazin, D. J. Singh, M. D. Johannes, M. H. Du,
{\it Phys. Rev. Lett.} {\bf 101}, 057003 (2008)
\bibitem{Umm09} G. A. Ummarino, M. Tortello, D. Daghero, R. S. Gonnelli,
{\it Phys. Rev.} {\bf B 80}, 172503 (2009)
\bibitem{Dag09} D. Daghero, M. Tortello, R. S. Gonnelli, V. A. Stepanov,
N. D. Zhigadlo, J. Karpinski, {\it Phys. Rev.} {\bf B 80}, 060502(R)
(2009)
\bibitem{Lee10} H.-S. Lee, M. Bartkowiak, J. S. Kim, H.-J.
Lee, {\it Phys. Rev.} {\bf B 82}, 104523 (2010)
\bibitem{Pra11} G. Prando, P. Carretta, R. De Renzi, S. Sanna, A.
Palenzona, M. Putti, M. Tropeano, {\it Phys. Rev.} {\bf B 83},
174514 (2011)
\bibitem{Bla94} G. Blatter, M. V. Feigel'man, V. B. Geshkenbein,
A. I. Larkin, V. M. Vinokur, {\it Rev. Mod. Phys.} {\bf 66}, 1125
(1994)
\bibitem{Koc89} R. H. Koch, V. Foglietti, W. J. Gallagher, G. Koren,
A. Gupta, M. P. A. Fisher, {\it Phys. Rev. Lett.} {\bf 63}, 1511
(1989)
\bibitem{Fis91} D. S. Fisher, M. P. A. Fisher, D. A. Huse, {\it Phys. Rev.}
{\bf B 43}, 130 (1991)
\bibitem{Che08} P. Cheng, H. Yang, Y. Jia, L. Fang, X. Zhu, G. Mu, H. H.
Wen, {\it Phys. Rev.} {\bf B 78}, 134508 (2008)
\bibitem{Sat08} T. Sato, S. Souma, K. Nakayama, K. Terashima, K. Sugawara,
T. Takahashi, Y. Kamihara, M. Hirano, H. Hosono, {\it Journ. Phys.
Soc. Jpn.} {\bf B 77}, 063708 (2008)
\bibitem{Mer09} T. Mertelj, V. V. Kabanov, C. Gadermaier, N. D. Zhigadlo,
S. Katrych, J. Karpinski, D. Mihailovic, {\it Phys. Rev. Lett.} {\bf
102}, 117002 (2009)
\bibitem{Gon09} R. S. Gonnelli, D. Daghero, M. Tortello, G. A. Ummarino,
V. A. Stepanov, J. S. Kim, and R. K. Kremer, {\it Phys. Rev.} {\bf B
79}, 184526 (2009)
\bibitem{Pal11} I. Pallecchi, M. Tropeano, C. Ferdeghini, G. Lamura,
A. Martinelli, A. Palenzona, M. Putti, {\it J. Supercond. Nov.
Magn.}, in press
\bibitem{Sko75} W. J. Skocpol, M. Tinkham, {\it Rep. Prog. Phys.} {\bf
B 38}, 1049 (1975)
\bibitem{Tin96} M. Tinkham, {\it Introduction to Superconductivity},
McGraw-Hill Book Co. (1996)
\bibitem{Lar05} A. Larkin, A. Varlamov, {\it Theory of Fluctuations in
Superconductors}, Oxford Science Publications (2005)
\bibitem{Las02a} A. Lascialfari, T. Mishonov, A. Rigamonti, P. Tedesco, A.
Varlamov, {\it Phys. Rev.} {\bf B 65}, 180501(R) (2002)
\bibitem{Las02b} A. Lascialfari, A. Rigamonti, L. Roman\'o, P. Tedesco, A.
Varlamov, D. Embriaco, {\it Phys. Rev.} {\bf B 65}, 144523 (2002)
\bibitem{Mur10} J. M. Murray, Z. Tesanovic, {\it Phys. Rev.
Lett.} {\bf 105}, 037006 (2010)
\bibitem{Wel11} U. Welp, C. Chaparro, A. E. Koshelev, W. K. Kwok, A.
Rydh, N. D. Zhigadlo, J. Karpinski, S. Weyeneth, {\it Phys. Rev.}
{\bf B 83}, 100513 (2011)
\bibitem{Mos11} J. Mosqueira, J. D. Dancausa, F. Vidal, S. Salem-Sugui Jr.,
A. D. Alvarenga, H.-Q. Luo, Z.-S. Wang, H.-H. Wen, {\it Phys. Rev.}
{\bf B 83}, 094519 (2011)
\bibitem{Sew01} A. Sewer, H. Beck, {\it Phys. Rev.} {\bf B 64}, 014510
(2001)
\bibitem{Rom03} L. Roman\'o, {\it Int. Journ. Mod. Phys.} {\bf
B 17}, 423 (2003)
\bibitem{Mar08} A. Martinelli, M. Ferretti, P. Manfrinetti, A. Palenzona,
M. Tropeano, M. R. Cimberle, C. Ferdeghini, R. Valle, C. Bernini, M.
Putti, A. S. Siri, {\it Supercond. Sci. Technol.} {\bf 21} 095017
(2008)
\bibitem{Mar09} S. Margadonna, Y. Takabayashi, M. T. McDonald, M.
Brunelli, G. Wu, R. H. Liu, X. H. Chen, K. Prassides, {\it Phys.
Rev.} {\bf B 79}, 014503 (2009)
\bibitem{Pra10} G. Prando, P. Carretta, A. Rigamonti, S. Sanna, A.
Palenzona, M. Putti, M. Tropeano, {\it Phys. Rev.} {\bf B 81},
100508(R) (2010)
\bibitem{Kos94} A. E. Koshelev, {\it Phys. Rev.} {\bf B 50}, 506 (1994)
\bibitem{Hub95} M. A. Hubbard, M. B. Salamon, B. W. Veal, {\it Physica}
{\bf C 259}, 309 (1995)
\bibitem{Mis00} T. Mishonov, E. Penev, {\it Int. Journ. Mod. Phys.}
{\bf B 14}, 3831 (2000)
\bibitem{Rom05} L. Roman\'o, A. Lascialfari, A. Rigamonti, I. Zucca,
{\it Phys. Rev. Lett.} {\bf 94}, 247001 (2005)
\bibitem{Ber08} E. Bernardi, A. Lascialfari, A. Rigamonti, L. Roman\'o,
{\it Phys. Rev.} {\bf B 77}, 064502 (2008)
\bibitem{Jun98} A. Junod, J.-Y. Genoud, G. Triscone, T. Schneider,
{\it Physica} {\bf C 294}, 115 (1998)
\bibitem{Ber06} E. Bernardi, A. Lascialfari, A. Rigamonti, L. Roman\'o,
V. Iannotti, G. Ausanio, C. Luponio, {\it Phys. Rev.} {\bf B 74},
134509 (2006)
\bibitem{Las03a} A. Lascialfari, A. Rigamonti, P. Tedesco, L.
Roman\'o, A. Varlamov, D. Embriaco, {\it Int. Journ. Mod. Phys.}
{\bf B 17}, 785 (2003)
\bibitem{Las03b} A. Lascialfari, A. Rigamonti, L. Roman\'o, A. A.
Varlamov, I. Zucca, {\it Phys. Rev.} {\bf B 68}, 100505(R) (2003)
\bibitem{Li10} L. Li, Y. Wang, S. Komiya, S. Ono, Y. Ando, G. D. Gu,
N. P. Ong, {\it Phys. Rev.} {\bf B 81}, 054510 (2010)
\bibitem{Ber10} E. Bernardi, A. Lascialfari, A. Rigamonti, L. Roman\'o,
M. Scavini, C. Oliva, {\it Phys. Rev.} {\bf B 81}, 064502 (2010)
\bibitem{Eme95} V. J. Emery, S. A. Kivelson, {\it Nature} {\bf 374}, 434
(1995)
\bibitem{Xu00} Z. A. Xu, N. P. Ong, Y. Wang, T. Kakeshita, S. Uchida,
{\it Nature} {\bf 406}, 486 (2000)
\bibitem{Igu01} I. Iguchi, T. Yamaguchi, A. Sugimoto,
{\it Nature} {\bf 412}, 420 (2001)
\bibitem{Lan02} K. M. Lang, V. Madhavan, J. E. Hoffman, E. W. Hudson,
H. Eisaki, S. Uchida, J. C. Davis, {\it Nature} {\bf 415}, 412
(2002)
\bibitem{Wan06} Y. Wang, L. Li, N. P. Ong, {\it Phys. Rev.} {\bf B 73},
024510 (2006)
\bibitem{Cab06} L. Cabo, F. Soto, M. Ruibal, J. Mosqueira, F. Vidal,
{\it Phys. Rev.} {\bf B 73}, 184520 (2006)
\bibitem{Rig05} A. Rigamonti, A. Lascialfari, L. Roman\'o, A. Varlamov,
I. Zucca, {\it J. Supercond. Incorp. Nov. Magn.},
DOI:10.1007/s10948-005-0077-z
\bibitem{Rya09} D. H. Ryan, J. M. Cadogan, C. Ritter, F. Canepa,
A. Palenzona, M. Putti, {\it Phys. Rev.} {\bf B 80}, 220503(R)
(2009)
\bibitem{Van32} J. H. Van Vleck, {\it The Theory of Electric and
Magnetic Susceptibilities}, Oxford University Press (1932)
\bibitem{Chi08} S. Chi, D. T. Adroja, T. Guidi, R. Bewley, S. Li,
J. Zhao, J.W. Lynn, C. M. Brown, Y. Qiu, G. F. Chen, J. L. Lou, N.
L. Wang, P. Dai, {\it Phys. Rev. Lett} {\bf 101} 217002 (2008)
\bibitem{Bak09} P. J. Baker, S. R. Giblin, F. L. Pratt, R. H. Liu,
G. Wu, X. H. Chen, M. J. Pitcher, D. R. Parker, S. J. Clarke, S. J.
Blundell, {\it New Journ. Phys.} {\bf 11} 025010 (2009)
\bibitem{Cim09} M. R. Cimberle, F. Canepa, M. Ferretti, A.
Martinelli, A. Palenzona, A. S. Siri, C. Tarantini, M. Tropeano, C.
Ferdeghini, {\it Journ. Magn. Magn. Mat.} {\bf 321} 3024 (2009)
\bibitem{Fan09} L. Fanfarillo, L. Benfatto, S. Caprara, C. Castellani,
M. Grilli, {\it Phys. Rev.} {\bf B 79}, 172508 (2009)
\bibitem{Pal09} I. Pallecchi, C. Fanciulli, M. Tropeano, A.
Palenzona, M. Ferretti, A. Malagoli, A. Martinelli, I. Sheikin, M.
Putti, C. Ferdeghini, {\it Phys. Rev.} {\bf B 79}, 104515 (2009)
\bibitem{Lee09} H.-S. Lee, M. Bartkowiak, J.-H. Park, J.-Y. Lee,
J.-Y. Kim, N.-H. Sung, B. K. Cho, C.-U. Jung, J. S. Kim, H.-J. Lee,
{\it Phys. Rev.} {\bf B 80}, 144512 (2009)
\bibitem{Sch93} T. Schneider, H. Keller, {\it Physica} {\bf
C 207}, 366 (1993)
\bibitem{Sch94} T. Schneider, H. Keller, {\it Int. Journ. Mod. Phys.}
{\bf B 8}, 487 (1994)
\bibitem{Tal11} J. L. Tallon, J. G. Storey, J. W. Loram, {\it Phys. Rev.} {\bf B
83}, 092502 (2011)
\end{references}
\end{document}